\documentclass[aps,twocolumn,secnumarabic,nobalancelastpage,amsmath,amssymb]{revtex4-1}

\usepackage[utf8]{inputenc}
\usepackage{geometry}
\usepackage{graphicx}
\usepackage{array}
\usepackage{slashed}
\usepackage{amsmath}
\usepackage{amsfonts}
\usepackage{amssymb}
\usepackage[cm]{fullpage}
\usepackage{epsfig}
\usepackage{changes}
\usepackage{comment}
\usepackage{hyperref}
\usepackage{tikz}
\usepackage{feynmp}
\usepackage{cancel}
\usepackage{mathrsfs}
\usepackage{dcolumn}
\usepackage{bm}
\usepackage{balance}
\usepackage{physics}
\usepackage{xspace}
\usepackage{xfrac}
\usepackage{tensor}
\usepackage{mathtools}
\usepackage{ragged2e}

\usepackage{floatrow}
\usepackage[caption=false]{subfig}
\usepackage[acronym]{glossaries}
\glsdisablehyper

\usepackage{cleveref} 
\definechangesauthor[name=Vincent,color=violet]{hzc}
\definechangesauthor[name=Ana,color=orange]{amr}

\DeclareCaptionJustification{justified}{\justifying}
\newcommand{\resetLabelSep}{\setlength{\labelsep}{5pt}}
\newcommand{\flushLabelSep}{\setlength{\labelsep}{-13pt}}

\newcommand{\bfw}{\mathbf{w}}
\newcommand{\bfz}{\mathbf{z}}

\newacronym{AdS}{AdS}{anti-de Sitter}
\newacronym{AdSCFT}{AdS/CFT}{anti-de Sitter/conformal field theory}
\newacronym{C}{C}{celestial}
\newacronym{CCFT}{CCFT}{celestial CFT}
\newacronym{CFT}{CFT}{conformal field theory}
\newacronym{CPW}{CPW}{conformal primary wavefunction}

\newcommand{\eg}{\textit{e.g.\@\xspace}}

\newcommand{\ie}{\textit{i.e.,}\ }

\newcommand{\CFT}{\mathrm{CFT}}

\newcommand{\cs}{\mathrm{CS}}
\newcommand{\csI}{\mathrm{CSI}}
\newcommand{\edge}{\mathrm{E}}
\newcommand{\edgeI}{\mathrm{EI}}

\newcommand{\ent}{\mathrm{ent}}
\newcommand{\gold}{\mathrm{G}}

\newcommand{\p}{\partial}

\newcommand{\eqReg}{\stackrel{\reg}{=}}
\newcommand{\eqIntegrated}{\stackrel{\reg}{\cong}}
\newcommand{\eqOnShell}{\mathrel{\hat{=}}}

\newcommand{\reals}{\mathbb{R}}

\newcommand{\volForm}{\epsilon}
\newcommand{\stepFunc}{\Theta}

\newcommand{\lap}{\Box}

\newcommand{\flatPlane}[1]{\reals^{#1}}

\newcommand{\minkX}{X}
\newcommand{\minku}{u}
\newcommand{\minkr}{r}

\newcommand{\minkzh}{z}

\newcommand{\lMilne}{L}
\newcommand{\rMilne}{R}
\newcommand{\milnet}{\tau}

\newcommand{\hyp}[1]{H_{#1}}

\newcommand{\fgr}{\rho} 
\NewDocumentCommand{\nullInfty}{o}{
  \IfNoValueTF{#1}
  {\mathscr{I}}
  {\mathscr{I}^{#1}}%
}

\newcommand{\spaceInfty}{i_0}

\newcommand{\nullUnit}{\hat{q}}

\newcommand{\polar}{\varepsilon}

\newcommand{\celPlane}{\flatPlane{2}}
\newcommand{\celMet}{\gamma}
\newcommand{\celVolForm}{\volForm^{(2)}}

\newcommand{\celLap}{\lap^{(2)}} 

\NewDocumentCommand{\celw}{o}{%
  \IfNoValueTF{#1}
  {\mathbf{w}}
  {w#1}%
}
\newcommand{\celwh}{w}

\DeclareMathOperator{\celG}{G}
\DeclareMathOperator{\celDeltaFunc}{\delta}

\newcommand{\celFreq}{\omega}

\newcommand{\invTemp}{\beta}
\newcommand{\reg}{\epsilon}
\newcommand{\act}{K} 
\newcommand{\actInt}{\mathcal{\act}}

\newcommand{\shad}{\tilde}

\newcommand{\inv}{\underline}

\def\bz{\bar{z} }

\begin{document}

\title{\Large{\bf Entanglement, Soft Modes, and Celestial Holography }}

\author{Hong Zhe Chen$^{*,\dagger}$, Robert Myers$^{*}$, and Ana-Maria
  Raclariu$^{*, \ddagger}$}

\affiliation{%
  $^*$Perimeter Institute for Theoretical Physics, 31 Caroline Street North,
  Waterloo, N2L 2Y5, Canada%
}

\affiliation{%
  $^\dagger$Department of Physics \& Astronomy, University of Waterloo,
  Waterloo, ON N2L 3G1, Canada%
}

\affiliation{%
  $^{\ddagger}$Institute for Theoretical Physics, University of Amsterdam,
  Science Park 904, Postbus 94485, 1090 GL Amsterdam, The Netherlands%
}

\begin{abstract}
  We evaluate the vacuum entanglement entropy across a cut of future null
  infinity for free Maxwell theory in four-dimensional Minkowski spacetime. The
  Weyl invariance of 4D Maxwell theory allows us to embed the Minkowski
  spacetime inside the Einstein static universe. The Minkowski vacuum can then
  be described as a thermofield double state on the (future) Milne wedges of the
  original and inverted Minkowski patches. We show that the soft mode
  contribution to entanglement entropy is due to correlations between asymptotic
  charges of these Milne wedges or, equivalently, nontrivial conformally soft
  (or edge) mode configurations at the entangling surface.
\end{abstract}

\maketitle

\nopagebreak


Entanglement is a distinguishing feature of quantum physics, shaping many
properties of complex interacting systems. The entanglement between any
subsystem $\rMilne$ and its complement can be quantified by the entanglement
entropy:
\begin{align}
  \label{eq:vN}
  S_{\mathrm{vN}}(\tensor[^\rMilne]{\rho}{})
  = - \tr(\tensor[^\rMilne]{\rho}{} \log \tensor[^\rMilne]{\rho}{})\,,
\end{align}
\ie{} the von Neumann entropy of the reduced density matrix
$\tensor[^\rMilne]{\rho}{}$ describing the state on $\rMilne$. Remarkably, the anti-de Sitter/conformal field theory (AdS/CFT) correspondence \cite{Maldacena:1997re}
equates the entanglement entropy of subregions in the boundary \gls{CFT} with a
generalized gravitational entropy in the \gls{AdS} bulk \cite{Ryu:2006ef,
  Ryu:2006bv, Lewkowycz:2013nqa, Faulkner:2013ana, Rangamani:2016dms}. This
discovery has sparked many exciting advances in our understanding of quantum
gravity using the tools of quantum information, including connections to quantum
error correction \cite{Almheiri:2014lwa,Pastawski:2015qua} and insights into the
black hole information paradox \cite{Almheiri:2019psf, Penington:2019npb,
  Almheiri:2019hni}.

Concurrently, gauge and gravity theories in asymptotically flat spacetimes were
shown to have a very rich infrared structure \cite{Strominger:2013lka,
  Strominger:2013jfa, He:2014laa, He:2014cra}. Here, the vacuum is
infinitely degenerate, with different vacua related by asymptotic symmetries or,
equivalently, the addition of soft particles \cite{He:2014cra, He:2014laa,
  Kapec:2014opa, He:2015zea}. Arising from these developments, celestial
holography, \eg{} \cite{Strominger:2017zoo, Raclariu:2021zjz,Pasterski:2021rjz},
proposes a duality between a $(3+1)$-dimensional asymptotically flat spacetime
and a two-dimensional \gls{CCFT}. 
Entanglement and the representation of bulk subregions in the new holographic framework have been unexplored so far.

\begin{figure}[h]
  \hfill \sidesubfloat[]{
    \includegraphics[scale=0.65]{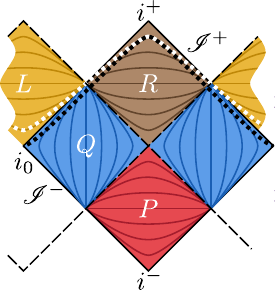}
    \label{fig:bulkSubregions}
  } \hfill \sidesubfloat[]{
    \includegraphics[scale=0.65]{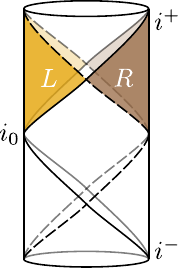}
    \label{fig:einsteinStaticUniverse}
  } \hfill\break
  \caption{\protect\subref{fig:bulkSubregions} Penrose diagram of Minkowski
    spacetime and associated inverted spacetime
    (delineated by solid and dashed diagonal lines, respectively). They overlap
    in the patch $Q$, while $P,R,L$ denote Milne patches.
    Minkowski (black) and Milne (white) Cauchy slices near $\mathscr{I}^+$ are
    drawn as dashed curves. \protect\subref{fig:einsteinStaticUniverse}
    Conformal image in the Einstein
    universe.}
	\label{fig:cauchySlices}
\end{figure}

In this letter we take essential steps in formulating a new entry in the  flat space holographic dictionary relating bulk subregions in 4D Minkowski spacetime and observables in 2D celestial
CFT.
It was noted in \cite{Kapec:2016aqd} that infrared effects may contribute
nontrivially to the entanglement entropy across a cut on future null infinity
$\mathscr{I}^+$. Pursuing this direction further, we examine the vacuum
entanglement across a cut on $\mathscr{I}^+$ for free Maxwell theory in
four-dimensional Minkowski spacetime. This entangling surface on $\mathscr{I}^+$
is defined by the future lightcone emanating from a point in the
Minkowski geometry. Following standard convention, we refer to the spacetime
region to the future of this lightcone as the future 
Milne patch -- see \cref{fig:bulkSubregions}. Hence, we are considering the
entanglement entropy \ref{eq:vN} for the mixed state on this region.

To this end, we leverage the Weyl invariance of 4D Maxwell theory to embed the
Minkowski spacetime inside the Einstein static universe $\reals\times S^3$. In
this context, a full Cauchy slice is formed by Cauchy slices for Milne wedges of
the original geometry ($\rMilne$) and its counterpart ($\lMilne$) in the
``inverted" Minkowski geometry produced with a conformal inversion of the
original spacetime -- see \cref{fig:cauchySlices}. We will show that the mixed
state $\tensor[^\rMilne]{\rho}{}$ can be purified as a thermofield double state
on $\lMilne$ and $\rMilne$. Along the way, we develop new methods to characterize conformal primary sectors of bulk subregions. A challenge in studying this entanglement is
then the treatment of soft modes. We attribute the associated entanglement to asymptotic charge fluctuations, and explain how spacetime entanglement could arise holographically from the celestial CFT. Due to the universality of soft physics, we expect similar methods to be useful
to describe the asymptotic physics of
spacetime fluctuations in (quantum) gravity \cite{Verlinde:2019xfb,Zurek:2020ukz,Verlinde:2022hhs}.



{\bf Preliminaries:} A key step in celestial holography is replacing plane-waves
by boost eigenfunctions, transforming as conformal primaries under the Lorentz
group \cite{Pasterski:2017kqt}. These conformal primary wavefunctions (CPWs) in free Maxwell theory in
(3+1)-dimensional Minkowski spacetime, with coordinates $\minkX$, are
\begin{align}
  \label{eq:spin-1-cpw}
  A^{\Delta, \pm}_{a; \mu}({\bf w}; X)
  &=  \frac{
    m^{\pm}_{a;\mu}({\bf w}; X)
    }{
    [-\nullUnit(\mathbf{w})\cdot \minkX_{\pm}]^{\Delta}},
  &
    \minkX_{\pm}^{\mu}
  &= \minkX^{\mu} \mp i\reg \hat{n}^{\mu}
    \;,
\end{align}
labelled by points ${\bf w}$ in celestial space, corresponding to a section of
$\mathscr{I}^{\pm}$ reached by null rays $\pm\nullUnit({\bf w})$. The
$m^{\pm}_{a;\mu}$ label photon polarizations, but their precise form will be
unimportant in the following \cite{Pasterski:2020pdk}\footnote{Our conventions
  are summarized in the Supplementary Material.}. With $\hat{n}$ obeying
$\hat{n}\cdot \nullUnit({\bf w})=1$, the $i\epsilon$ provides a prescription to
cross the $\hat{q} \cdot X = 0$ surface.

The modes \labelcref{eq:spin-1-cpw} have eigenvalues $\Delta$ under boosts
towards $\hat{q}(\mathbf{w})$, dual to dilations about $\mathbf{w}$ in celestial
space. These $+$/$-$ modes form a basis of solutions with
positive/negative-definite frequencies with respect to Minkowski time, provided
$\Delta = 1 + i\lambda, ~ \lambda \in \mathbb{R}$. Here, $\lambda$ is also a
frequency with respect to Milne time $\tau$:
\begin{align}
  \partial_\milnet
  A^{1+i\lambda}_a
  &\eqReg -i\lambda A^{1+i\lambda}_a\,,
  &
    \milnet
  &=
    \frac{1}{2}
    \log(-\minkX^2),
    \label{eq:milnet}
\end{align}
where $\eqReg$ denotes equality when $\reg\to 0$. The $\Delta = 1$ ($\lambda=0$)
modes yield the Goldstone wavefunctions $A^{\gold}_a({\bf w}, X)$, while their
canonically conjugate conformally soft partners $ A^{\cs}_a({\bf w}, X)$ (which
also have $\Delta = 1$) were constructed in \cite{Donnay:2018neh} -- see also
\cite{Note1}.

The asymptotic symmetries of pure Maxwell theory are generated by asymptotic
charges \cite{He:2014cra}\footnote{Asymptotic charge conservation states that
  \cref{eq:asymptotic-charge} is independent of the choice of $\pm$.}
\begin{align}
  \label{eq:asymptotic-charge}
  Q[\alpha]
  &= \int_{\mathscr{I}^{\pm}_{\mp}} \alpha * F
    \eqOnShell \int_{\mathscr{I}^{\pm}} d\alpha \wedge * F
    \;,
\end{align}
where $\mathscr{I}^{+}_{-}$ ($\mathscr{I}^-_+$) is the past (future) boundary of
$\mathscr{I}^+$ ($\mathscr{I}^-$) and the last equality holds on-shell. Turning
on $A^\cs$ changes the values of the charges
\eqref{eq:asymptotic-charge}. In contrast, the global charge obtained with
constant $\alpha$ vanishes in free Maxwell theory.

From CPWs, one constructs operators
\begin{align}
  \mathcal{O}^{\Delta,\pm}_a(\mathbf{w})
  &= -i \langle
    A^{\Delta,\pm}_a(\mathbf{w}),
    A
    \rangle
    =\mathcal{O}^{\Delta^*,\mp}_{\bar{a}}(\mathbf{w})^\dagger
    \;.
    \label{eq:confPrimOp}
\end{align}
Here $ \langle \cdot, \cdot \rangle$ denotes the spacetime inner product among spin-1 conformal primary wavefunctions \cite{Donnay:2018neh}.
Replacing $ A^{\Delta}$ by $A^{\gold}, A^{\cs}$ moreover defines the
(conformally) soft operators $ \mathcal{Q}_a(\mathbf{w}) =
Q[\alpha^{\gold}],\,\mathcal{S}_a(\mathbf{w})$ respectively. These can be
regarded as operators in the CCFT, exciting CPWs in the Minkowski
bulk with $S$-matrix elements encoded by CCFT correlation functions.

\section{Beyond the Minkowski patch}
\label{sec:beyond-Mink}

To prepare our entanglement calculations, we employ the Weyl invariance of 4D
Maxwell theory to extend the CPWs \eqref{eq:spin-1-cpw} beyond the Minkowski patch. We first consider
an inverted Minkowski patch covered by coordinates $\underline{X}^{\mu}$ -- see
\cref{fig:cauchySlices}. In the overlap $\underline{X}^2, X^2 >0$ with the
original Minkowski patch, an inversion $\underline{X}^{\mu} =
\frac{X^{\mu}}{X^2}$ and Weyl transformation relate the two geometries. Given
the Weyl invariance of the 4D Maxwell theory, the gauge field $\underline{A}$ in
the inverted patch is simply related to the original $A$ by
\begin{align}
  \underline{A}_{\mu}(\underline{X})
  = \frac{\p X^{\nu}}{\p \underline{X}^{\mu}} A_{\nu}(X)\;, \qquad X^2 > 0\;.
  \label{eq:inversionOfVectorField}
\end{align}
The inverted CPWs
\labelcref{eq:spin-1-cpw} are found to be proportional to shadow wavefunctions
\begin{align}
  \underline{A}^{\Delta,\pm}_{a,\mu}({\bf w}; \inv{\minkX})
  & \eqReg  e^{\pm i\pi (\Delta -1)}
    \tilde{A}^{\Delta,\pm}_{a, \mu}({\bf w};\inv{\minkX})\;.
    \label{eq:inversion}
\end{align}
 The shadow wavefunctions $\tilde{A}^{\Delta, \pm}_a$  \cite{Pasterski:2017kqt} yield an
alternate set of CPWs. We review their defining properties in \cite{Note1}.
The
$\reg\to0$ limit above ensures \cref{eq:inversion} crosses $\nullUnit\cdot
X=0=\nullUnit\cdot\underline{X}$ in the overlap region consistently with the
original mode \labelcref{eq:spin-1-cpw}. While \cref{eq:inversionOfVectorField}
only applies in the overlap of the two Minkowski patches, the result
\labelcref{eq:inversion} evolves to a solution over the full inverted patch.

Note that inversions and shadow transforms preserve the space of solutions and,
moreover, both $\underline{A}^{\Delta}$ and $\tilde{A}^{\Delta}$ have opposite
Milne frequency with respect to
$A^{\Delta}$. Therefore, a proportionality
\labelcref{eq:inversion} between the former modes is unsurprising. We expect a similar relation
between shadows and inversions to be a general property of Weyl-invariant
theories -- see also \cite{Brown:2022miw,Jorstad:2023ajr}.


{\bf Complementary Milne patches:} The two Minkowski spacetimes are also conformally mapped to
  the Einstein universe
$\reals\times S^3$ as in \cref{fig:einsteinStaticUniverse} \cite{Hawking:1973uf}. The future null Minkowski boundary
$\mathscr{I}^+$ (plus $\spaceInfty$) is a Cauchy surface for the Einstein
 universe. Another Cauchy surface is given by the union
of the Cauchy surfaces for the (future) Milne patches of the original and
inverted Minkowski patches, plus the surface (on $\mathscr{I}^+$) between them.
We denote these Milne patches as left ($\lMilne$) and right ($\rMilne$),
respectively. Hence, a state on $\mathscr{I}^+$ should admit a decomposition in
terms of $\lMilne,\rMilne$ states, as illustrated in
\cref{fig:cauchySlices}. This will allow us to evaluate the entanglement across the cut on $\mathscr{I}^+$
connecting the $\lMilne,\rMilne$ patches.

We proceed by constructing CPWs associated with the Milne patches. The modes
\eqref{eq:spin-1-cpw} can be decomposed into ${}^L\tilde{A}^{\Delta},
{}^RA^{\Delta}$ with initial data supported respectively on $\lMilne$ and
$\rMilne$ Cauchy surfaces
\begin{align}
  \label{eq:LR-dec}
  A^{\Delta,\pm}
  &\eqReg e^{\pm i\pi (\Delta -1)} {}^L\tilde{A}^{\Delta}
    +  {}^RA^{\Delta}\;, \quad \Delta \notin  \mathbb{Z}\;.
\end{align}
Here, $A^{\Delta, \pm}$ agrees with ${}^RA^{\Delta}$ inside $\rMilne$,
 and with $e^{\pm i\pi (\Delta
  -1)}\, {}^L\tilde{A}^{\Delta}$ inside $\lMilne$, upon extending
\cref{eq:inversion} to the full inverted Minkowski patch. Hence, for the
principal series (\ie{} $\Delta = 1 + i\lambda, ~ \lambda \in \mathbb{R}$), the
non-soft (\ie{} $\lambda\ne0$) modes ${}^{\lMilne}\tilde{A}^{\Delta},
{}^{\rMilne}A^{\Delta}$ are defined by \cref{eq:LR-dec}.

An alternate definition of the $\lMilne,\rMilne$ modes retains the $\epsilon$
regulator to make \cref{eq:LR-dec} an exact equality. Then
${}^{\lMilne}\tilde{A}^{\Delta}$ and $ {}^{\rMilne}A^{\Delta}$ are smooth
finite 
energy solutions. For instance, their Minkowski inner products can be evaluated
using the standard inner products among Minkowski CPWs
\cite{Donnay:2020guq,Pasterski:2017kqt} -- see also \cite{Note1}. Since
$\mathscr{I}^+$ is a Minkowski Cauchy surface and splits into complementary
$\lMilne$ and $\rMilne$ Cauchy surfaces, the inner product decomposes as
\begin{align}
  \langle \cdot, \cdot \rangle
  \eqIntegrated  {}^{\lMilne}\langle \cdot, \cdot \rangle
  +  {}^{\rMilne}\langle \cdot, \cdot \rangle\,.
  \label{eq:LR-soft-ip}
\end{align}
Thus, the Minkowski inner products of ${}^{\lMilne}\tilde{A}^{\Delta}$ and $
{}^{\rMilne}A^{\Delta}$ 
yield inner products ${}^{\lMilne}\langle \cdot, \cdot \rangle, {}^{\rMilne}\langle \cdot,
\cdot \rangle$ in the Milne
theories.

{\bf Conformally soft modes revisited:}
Here, we reveal $A^{\cs}$ as configurations sourced by charged particles in the
Einstein universe 
beyond the original Minkowski patch.

While $A^\gold$ has
vanishing field-strength, $A^\cs$ possesses electric fields localized on the $\hat{q}\cdot X =
0$ plane and the $X^{2} = 0$ lightcone. An
expansion near $\mathscr{I}^{\pm}$ reveals that the former leads to a nontrivial
electric field $F_{ru}$ near $i^0$ ``sourced'' by the radiative modes $F_{uz}$
\cite{Kapec:2017tkm}. Hence turning on $A^{\cs}$ yields a nontrivial asymptotic
charge \eqref{eq:asymptotic-charge}
\begin{align}
  \label{eq:lgc}
  Q^{\pm}[\alpha, A^{\cs}] = 4\pi\, d\alpha
  \;.
\end{align}

To understand this better, we introduce
\begin{equation}
  \begin{split}
    A^{\csI}({\bf w_1},{\bf w_2}; X)
    &\equiv \int_{{\bf w}_1}^{{\bf w}_2} A^{\cs}(\bfw ;X)
      \label{eq:lettuce}
  \end{split}
\end{equation}
(which is path-independent, because $A^\cs$ is exact on celestial space).
The field strength of $A^{\csI}$ is supported on three shockwaves:
Two planar shockwaves along $\nullUnit_1\cdot X =0$ and $\nullUnit_2\cdot
X=0$ carry electric field lines in from infinity, which then transfer to a
spherical shockwave at $X^2=0$. 
%

\begin{figure}[h!]
  \flushLabelSep \sidesubfloat[]{
    \includegraphics[scale=0.6]{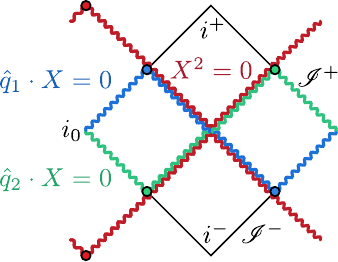}
    \label{fig:croissant}
  } \resetLabelSep 
  \hspace{1.5em}\sidesubfloat[]{
    \includegraphics[scale=0.6]{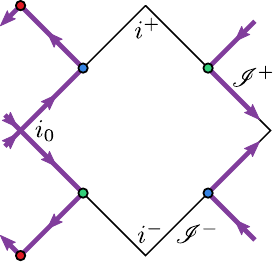}
    \label{fig:bun}
  }
  \caption{\protect\subref{fig:croissant}  Shockwaves of $A^{\csI}$
    extended beyond the Min-kowski patch. \protect\subref{fig:bun} Charged
    particles sourcing the shockwaves. }
  \label{fig:bread}
\end{figure}


\Cref{fig:croissant} draws the shockwaves of $A^{\csI}$ in the Einstein
universe, where a new interpretation of the charges \eqref{eq:lgc} emerges.
From this perspective, the same field configuration arises as the
Lienard-Wiechert fields of two oppositely charged particles,
as shown in \cref{fig:bun}.
Moreover, their trajectories contain kinks that act as sources and sinks for the
shockwaves. We emphasize that the shockwaves propagate in the original Minkowski
patch, but the charges sourcing them do not. Still, the effect of these charges
manifests quite physically in the Minkowski theory, giving precisely
\footnote{One may also consider time-like sources passing through $i^0$,
  contributing additionally to $Q^{\pm}$ through an $\hyp{3}$ bulk-to-boundary
  propagator.} the asymptotic charges \labelcref{eq:asymptotic-charge} as a
mathematical distribution $\alpha \mapsto Q^{\pm}[\alpha]$. For example,
\cref{eq:lgc} matches the dipole source of $A^\cs$ in \cref{eq:lettuce}.


{\bf Soft modes in Milne patches:}
By analogy with $A^\gold,A^\cs$, we expect pairs of canonically conjugate soft
($\Delta=1$) modes supported in the $\lMilne$ and $\rMilne$ patches. As we
show in \cite{Chen:2024kuq}, simply taking the $\Delta \rightarrow 1$ limit of
${}^{\lMilne}\tilde{A}^{\Delta}, {}^{\rMilne}A^{\Delta}$ violates matching
conditions at $i^0$ \cite{Strominger:2013jfa,Satishchandran:2019pyc}.

A decomposition respecting the matching conditions can be found by
considering the extension of $A^\gold,A^\cs$ outside the Minkowski
patch discussed above, with charges
running in between $\lMilne,\rMilne$:
\begin{align}
  A^{\gold}
  &= {}^{\lMilne}A^{\gold} +  {}^{\rMilne}A^{\gold},
  \quad
  A^{\cs} =  \, {}^{\lMilne}A^{\gold}
            + {}^{\lMilne}A^{\edge}
            + {}^{\rMilne}A^{\edge}
            \;.
            \label{eq:ASoftAsLR}
\end{align}
Here, we have introduced the ``edge'' modes ${}^LA^{\edge}, {}^RA^{\edge}$ which
are localized to an $\epsilon$-regulated shockwave as shown in
\cref{fig:AcsOnCauchySlices} and ensure smoothness of $A^{\cs}$ in
\cref{eq:ASoftAsLR} at finite $\reg$. As $\epsilon \rightarrow 0$, the
shockwave retreats from the Milne patches, leaving
\begin{align}
  {}^{\lMilne}A^{\gold}
  &= \lim_{\epsilon \rightarrow 0} A^{\cs},
    \quad {}^{\rMilne}A^{\gold}
    = \lim_{\epsilon \rightarrow 0}\left(A^{\gold} -   A^{\cs}\right)
    \;,
    \label{eq:LR-Mink-sourced}
\end{align}
which are pure gauge in their respective Milne interiors.

\begin{figure}[h!]
    \includegraphics[scale=0.6]{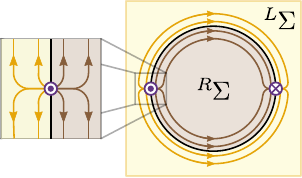}
    \label{fig:AcsOnCauchySlices}
 \setlength{\labelsep}{-5pt} 
  \caption{\label{fig:AcsOnCauchySlices} Electric fields of edge modes
    ${}^{\rMilne}A^{\edgeI}$ (brown) and ${}^{\lMilne}A^{\edgeI}$ (yellow) are part of the $\reg$-regulated $\minkX^2=0$ shockwave of
    $A^{\csI}$.
    Side panel shows how the
    shockwave contributes to the normal component
    of the electric field at $\partial{}^{\lMilne}\Sigma$ and    $\partial{}^{\rMilne}\Sigma$. }
\end{figure}

Eq.~\eqref{eq:LR-soft-ip} implies that these $\lMilne,\rMilne$ Goldstone and edge
modes are canonically conjugate. Note that only two independent combinations of
the four soft modes are permitted in the Minkowski theory, namely $A^{\rm CS}$
and $A^{\gold}$.

The Goldstone and edge modes augment the phase space of \eg{} the $\rMilne$
Milne patch with soft degrees of freedom ${}^{\rMilne}\mathcal{Q},
{}^{\rMilne}\mathcal{S}$ defined by the inner product of $A$ with
${}^{\rMilne}A^{\gold}$ and ${}^{\rMilne}A^{\edge}$ respectively, as in
\cref{eq:confPrimOp}. Paralleling \cref{eq:asymptotic-charge}, the
$\rMilne$ asymptotic charge reads
\begin{align}
  {}^{\rMilne}Q[\alpha]
  &= \int_{\partial {}^{\rMilne}\Sigma} \alpha * F
    \;,
    \label{eq:RCharge}
\end{align}
where ${}^\rMilne\Sigma$ is an $\rMilne$ Cauchy surface. By
\cref{eq:ASoftAsLR},
\begin{align}
  \mathcal{Q}
  &= {}^{\lMilne}\mathcal{Q}
    + {}^{\rMilne}\mathcal{Q} \;,
  &
    \mathcal{S}
  &= {}^{\lMilne}\mathcal{Q}
    + {}^{\lMilne}\mathcal{S}
    + {}^{\rMilne}\mathcal{S}
    \; .
    \label{eq:Q-S}
\end{align}

\section{Entanglement}
\label{sec:entanglement}

Let us now study how states in the Minkowski and Einstein Hilbert spaces embed
into the product space of the $L$ and $R$ Milne patches. We start by showing that
the common vacuum of the former is thermal with respect to the latter. After
analyzing constraints on the soft modes, we evaluate their contribution to the entanglement between the $\lMilne$ and $\rMilne$ patches.

{\bf Relation between vacua:} 
Weyl invariance of Maxwell theory allows us to prepare the vacuum
state $\ket{0}$ of the Minkowski and Einstein spacetimes with a Euclidean path
integral over an $S^4$ hemisphere. As described in the appendix, the same path integral prepares an entangled thermal state from the Milne
perspective. As a result, the vacuum state $\ket{0}$ is identified with a thermofield double state on $\lMilne$ and $\rMilne$.
An analogous path integral argument leads to the thermality of Rindler wedges and could alternatively be deduced from an analysis of the Bogoliubov transformations relating the Minkowski and Milne modes. In particular, demanding that 
$\ket{0}$ and the Milne vacua
$\tensor[^{\lMilne/\rMilne}]{\ket{0}}{}$ be annihilated by the respective
annihilation operators, one finds
\begin{align}
  \ket{0}
  &\propto e^{- \actInt^+}
    \tensor[^{\lMilne}]{\ket{0}}{}
    \tensor[^{\rMilne}]{\ket{0}}{},
    \label{eq:relationBetweenVacua}
\end{align}
where the entangling operator is given by
\begin{align}
  \actInt^+
  &= -\int_0^\infty d\lambda\,
    e^{-\pi\lambda}
    \int \celVolForm\,
    (\tensor[^\lMilne]{\shad{a}}{}^{1+i\lambda})^\dagger
    \cdot
    (\tensor[^\rMilne]{a}{^{1-i\lambda}})^\dagger
    \label{eq:horse}
\end{align}
in terms of Milne creation operators. Here, $\celVolForm$ is the volume form
over celestial space \cite{Note1}.

Turning to the soft modes, rather than relying on the $\lambda\to 0$ limit of
our analysis above, we instead study the constraints \cite{Donnelly:2016auv}
that entangle the edge modes \cite{Donnelly:2015hxa,Donnelly:2014fua}. (See
related results in \cite{Huang:2014pfa,Ball:2024hqe}.)

{\bf Soft constraints:}
In gauge theory, upon stitching together two complementary regions
${}^L\Sigma,{}^R\Sigma$ of a Cauchy slice, the physical Hilbert space
$\mathcal{H}$ is not simply the product of the $\lMilne,\rMilne$ Hilbert spaces
${}^L\mathcal{H} \otimes {}^R \mathcal{H}$. Rather, the admissible physical
states satisfy constraints \cite{Donnelly:2016auv}.

For example, Gauss's law at the entangling surface $\p {}^L\Sigma = -\p {}^R
\Sigma$ requires continuity of the normal component of the electric field
\cite{Donnelly:2016auv}, $\mathcal{Q}^{\ent\,\prime} \equiv {}^L \mathcal{Q} +
{}^R \mathcal{Q} = 0$.
The Hilbert space $\mathcal{H}$ of the theory covering the full Cauchy surface
is then the kernel $ \mathcal{H}
  = \ker \mathcal{Q}^{\ent\,\prime}
    \subset {}^L \mathcal{H} \otimes {}^R \mathcal{H}$.
This relates the Hilbert spaces of the
Einstein universe and $\lMilne,\rMilne$ Milne patches.

However, $\mathcal{Q}^{\ent\,\prime}$ is precisely the asymptotic charge
$\mathcal{Q}$ in \cref{eq:Q-S}, which is nontrivial in the Minkowski Hilbert
space. Instead, an appropriate constraint annihilating this Hilbert
space can be selected by finding a linear combination of $\lMilne, \rMilne$ soft
operators that commutes with the Minkowski operators $\mathcal{Q}$ and
$\mathcal{S}$:
\begin{align}
  \mathcal{Q}^{\ent}
  &=
    {}^{\lMilne}\mathcal{Q}
    + {}^{\rMilne}\mathcal{Q}
    +2\left(
    {}^{\lMilne}\mathcal{S}
    - {}^{\rMilne}\mathcal{S}
    \right)
    \;.
    \label{eq:entOpMink}
\end{align}
The extra terms account for the sources in \cref{fig:bun}, which violate Gauss's
law. Due to mixing with hard modes, $\mathcal{Q}^\ent$ in fact only annihilates
the Minkowski Hilbert space in the $\reg\to 0$ limit, as discussed later.

Because the vacuum $|0\rangle$ is shared by the Einstein and Minkowski theories,
it is annihilated by $\mathcal{Q}$ and $\mathcal{Q}^\ent$ (with $\reg\to 0$).
The Minkowski Hilbert space also contains states $|q\rangle$ with asymptotic
charges $Q[\alpha] = \int \celVolForm\, q\, \alpha$ parametrized by a celestial
scalar function $q(\bfw)$ \footnote{The mean of $q$ is required to vanish,
  corresponding to a vanishing global charge.}. These $|q\rangle$ carry a
background
\begin{align}
  A^{\cs}[q]
  &=
    \frac{1}{4\pi} \int \celVolForm(\mathbf{w})\,
    q(\mathbf{w})\,
    A^{\csI} (\mathbf{w},\infty)\,,
    \label{eq:Acsq}
\end{align}
produced by dressing \cite{Arkani-Hamed:2020gyp}
\begin{align}
  |q\rangle
  &= e^{i \mathcal{S}[q]} |0 \rangle
    \;,
  &
    \mathcal{S}[q]
  &= -i\langle A^\cs[q], A \rangle
    \;.
    \label{eq:softStateMink}
\end{align}
Since $\mathcal{S}[q]$ is linear in $A^\cs_a$, it is also linear in
$\mathcal{S}_a$. Hence, it can be shown using \cref{eq:Q-S} that
$[\mathcal{S}[q],\mathcal{Q}^{\ent}]=0$, so $|q\rangle$ satisfies the Minkowski
constraint.

{\bf Entanglement of edge modes:}
Let us now consider the $R$ reduced density matrices of the states $|q\rangle$
\begin{align}
  \tensor[^{\rMilne}]{\rho}{}[q]
  \equiv \tr_{\tensor[^L]{\mathcal{H}}{}}
  \ket{q}\bra{q}
  = e^{i\, {}^{\rMilne}\mathcal{S}[q]/2} \ 
  {}^{\rMilne}\rho[0] \, e^{-i\, {}^{\rMilne}\mathcal{S}[q]/2}.
\end{align}
Hence, ${}^{\rMilne}\rho[q]$ and ${}^{\rMilne}\rho[0]$ are unitarily related
because $ \mathcal{S}[q] = \frac{1}{2}\tensor[^R]{\mathcal{S}}{}[q] +
\tensor[^L]{\left(\cdots \right)}{}$
 \cite{Chen:2024kuq}.
Here, $\tensor[^{R}]{\mathcal{S}}{}[q]$ is defined in analogy to
\cref{eq:softStateMink}, using the $\rMilne$ inner
product with
\begin{align}
  \tensor[^\rMilne]{A}{^\edge}[q]
  &=
    \frac{1}{2\pi} \int \celVolForm(\mathbf{w})\,
    q(\mathbf{w})\,
    {}^\rMilne A^{\edgeI} (\mathbf{w},\infty)
    \;,
    \label{eq:AEq}
\end{align}
where ${}^\rMilne A^{\edgeI}$ is constructed from ${}^\rMilne A^{\edge}$ in the
same manner as \cref{eq:lettuce}. By design, $\tensor[^\rMilne]{A}{^\edge}[q]$
has Milne asymptotic charge $\tensor[^\rMilne]{Q}{}[\alpha] = \int \celVolForm\,
q\, \alpha$. Thus, all ${}^{\rMilne} \rho[q]$ share the same spectrum and von
Neumann entropy \eqref{eq:vN}, being merely dressed by different edge mode
backgrounds.

Therefore, we focus on the entanglement of $|0\rangle$, which yields ${}^R
\rho[0]$. Because
\begin{align}
  [\tensor[^\rMilne]{\mathcal{Q}}{}, \tensor[^\rMilne]{\rho[0]}{}]
  &=
    -\tr_{\tensor[^\lMilne]{\mathcal{H}}{}}
    [\tensor[^\lMilne]{\mathcal{Q}}{}, \ket{0}\bra{0}]
    =0
    \;,
    \label{eq:QRrhoCommutation}
\end{align}
${}^{\rMilne}\rho[0]$ admits a decomposition into blocks
$\tensor[^R]{\rho}{}[0, q]$ of definite $^{R}\mathcal{Q}$ (or equivalently
$\tensor[^\rMilne]{Q}{}$) \cite{Donnelly:2015hxa,Donnelly:2014fua}
\begin{align}
  \tensor[^{\rMilne}]{\rho}{}[0]
  &= \int \mathcal{E}[q]\, p[q]\,
    \tensor[^\rMilne]{\rho}{}[0,q]\; ,
    \label{eq:dnsMatBlockDecomposition}
  \\
  \tensor[^\rMilne]{Q}{}[\alpha]\,
  \tensor[^\rMilne]{\rho}{}[0,q]
  &=
    \tensor[^\rMilne]{\rho}{}[0,q]\,
    \tensor[^\rMilne]{Q}{}[\alpha]
    =
    \left(
    \int \celVolForm \alpha\, q
    \right)
    \tensor[^\rMilne]{\rho}{}[0,q]
    \;,\nonumber
\end{align}
where $\mathcal{E}[q]$ and $p[q]$ are a measure and probability distribution
over the functions $q$. Further, $\tr\tensor[^R]{\rho}{}[0, q]=1$.

Isolating blocks $\tensor[^R]{\rho}{}[0, q]$ by fixing the Milne asymptotic
charge in a path integral representation of $\tensor[^{\rMilne}]{\rho}{}[0]$, we
find the blocks differ in their edge mode background \labelcref{eq:AEq}, but
share identical quantum fluctuations (due to the theory being free)
\cite{Donnelly:2015hxa,Donnelly:2014fua}. This leads to the unitary relation
\cite{Chen:2024kuq}
\begin{align}
  \tensor[^R]{\rho}{}[0, q]
  &= e^{i\tensor[^R]{\mathcal{S}}{}[q]} \ \tensor[^R]{\rho}{}[0, 0]
    \,e^{-i\tensor[^R]{\mathcal{S}}{}[q]}
    \;.
    \label{eq:blocksUnitaryRelation}
\end{align}
Consequently, the von Neumann entropy of \cref{eq:dnsMatBlockDecomposition}
decomposes into two independent pieces \cite{Donnelly:2014fua,Donnelly:2015hxa}
\begin{align}
  S_{\rm vN}\left(\tensor[^{R}]{\rho}{}[0] \right)
  &= S_{\rm Sh}(p) + S_{\rm vN}\left(\tensor[^{R}]{\rho}{}[0, 0] \right)
    \;,\label{EE55}
\end{align}
where the Shannon entropy $S_{\rm Sh}(p)$ of $p[q]$ is identified
as the edge mode contribution \footnote{R\'enyi entropies decompose similarly.}. In contrast to \cite{Donnelly:2014fua,Donnelly:2015hxa} the edge modes are identified with conformally soft modes \eqref{eq:lettuce}. We demonstrate in \cite{Chen:2024kuq} that these differ from the static modes 
in \cite{Donnelly:2014fua,Donnelly:2015hxa} by a gauge transformation. 

Further, our path integral analysis identifies the factor $p[q] \propto \exp[
-\mathcal{I}_{2\pi}[\tensor[^R]{A}{^\edge}[q]] ]$ due to the additive
contribution made by the background shift \labelcref{eq:AEq} to the Euclidean
action $\mathcal{I}_{2\pi}$ \footnote{${}^R \rho[0]$ is prepared with two copies
  of the Euclidean path integral in \cref{fig:pathIntegralMilne}, extending the
  range of $i\milnet$ to $2\pi$, hence the subscript on the action
  \(\mathcal{I}_{2\pi}\). The action \(\mathcal{I}_{\beta}\) includes both the
  usual Euclidean Maxwell action \(I_{\beta}\) outside a spacetime neighbourhood
  of the entangling surface and a surface term on the boundary of said
  neighbourhood. The surface term is necessary for a good variational principle
  when fixing \(q\) and its inclusion results in the total edge mode action
  \labelcref{eq:act-reg} being related to \(I_\beta\) by a sign:
  \(\mathcal{I}_{\beta}[\tensor[^\rMilne]{A}{^{\edge}}[q]]=-I_{\beta}[\tensor[^\rMilne]{A}{^{\edge}}[q]]\).
  To get the right sign in \cref{eq:act-reg}, we also work consistently with the
  Lorentzian definition of the electric field. For more details, see
  \cite{Chen:2024kuq}. The aforementioned \(I_{\beta}\) is the same as in
  \cite{Chen:2024kuq}.}. Explicitly \footnote{In \cref{eq:act-reg}, we have switched
  to a different $\reg$ regulator from that inherited from $A^\cs$ in
  \cref{eq:ASoftAsLR}. Following \cite{Donnelly:2014fua,Donnelly:2015hxa},
  $\tensor[^\rMilne]{A}{^{\edge}}$ is defined to be an \emph{exactly}
  Milne-static solution with electric field lines penetrating
  $\partial\tensor[^\rMilne]{\Sigma}{}$ as shown in
  \cref{fig:AcsOnCauchySlices}, except now $\partial\tensor[^\rMilne]{\Sigma}{}$
  is placed at a cutoff $\reg^2$ in the Fefferman-Graham coordinate $\rho=2u/r$
  along the Euclidean \gls{AdS} $\tensor[^\rMilne]{\Sigma}{}$. When $\reg\to 0$,
  the electric field lines bunch up to $\fgr=\reg^2\to 0$, similar to
  \cref{fig:AcsOnCauchySlices}. This static $\tensor[^\rMilne]{A}{^{\edge}}$
  agrees with $A^{\log}$ (with the old $\reg$ removed) in $R$ up to rescaling
  and a gauge transformation \cite{Chen:2024kuq}.},
\begin{align}
  \mathcal{I}_{\beta}[\tensor[^\rMilne]{A}{^\edge}[q]]
  &= \frac{-\invTemp}{2\log(\sfrac{1}{\reg})}
    \int \epsilon^{(2)} \,q\, (\celLap)^{-1}\,q
    + \order{\log^{-2}(\sfrac{1}{\reg})}
    \label{eq:act-reg}
\end{align}
where $(\celLap)^{-1}$ denotes convolution with the celestial Green's function
\cite{Note1}. The similarity to the edge mode action of
\cite{Donnelly:2014fua,Donnelly:2015hxa}, despite their focus on the spacetime
interior, is natural from the Einstein universe perspective.

\section{Discussion}
\label{sec:discussion}

Let us consider now 
the holographic interpretation of the entanglement studied
in this paper. Emulating \cref{eq:confPrimOp} using the Milne modes
\eqref{eq:LR-dec} and inner products \eqref{eq:LR-soft-ip}, we construct
operators
${}^\lMilne\shad{\mathcal{O}}^\Delta_a,{}^\rMilne{\mathcal{O}}^\Delta_a$.
In the $\eqIntegrated$ sense, ${}^\rMilne{\mathcal{O}}^\Delta_a$ is proportional
to annihilation and creation operators,
$\tensor[^\rMilne]{a}{}^{1-i\lambda}_a(\mathbf{w})$,
$\tensor[^\rMilne]{a}{}^{1-i\lambda}_{\bar{a}}(\mathbf{w})^\dagger$, for
$\Delta=1+i\lambda$ and $1-i\lambda$ (with $\lambda>0$), respectively. These
appear in a mode expansion of the field operator, multiplying the positive and
negative Milne frequency modes $({}^\rMilne{A}^{1+i\lambda})^a(\mathbf{w})$ and
$({}^\rMilne{A}^{1-i\lambda})^a(\mathbf{w})$, respectively. Similarly,
${}^\lMilne\shad{\mathcal{O}}^\Delta_a({\bf w})$ is proportional to
${}^\lMilne{\shad{a}}^{1+i\lambda}_a({\bf w})$ and
${}^\lMilne{\shad{a}}^{1+i\lambda}_{\bar{a}}({\bf w})^\dagger$. Holographically,
${}^\lMilne\shad{\mathcal{O}}^\Delta_a,{}^\rMilne{\mathcal{O}}^\Delta_a$ are seen as 
conformal primaries in two sectors
$\tensor[^\lMilne]{\CFT}{},\tensor[^\rMilne]{\CFT}{}$ dual to the respective bulk Milne
theories.

However, these sectors of the \gls{CCFT}
are not independent. To see this, we
can express the entangling operator \labelcref{eq:horse}
as a coupling between ${}^\lMilne\shad{\mathcal{O}}^\Delta$ and ${}^\rMilne{\mathcal{O}}^\Delta$. 
For example, amplitudes in the $\ket{0}$ state are evaluated by
\gls{CFT} correlation functions in the presence of this interaction:
\begin{align}
  \bra{0}
  \bullet
  \ket{0}
  &= \langle
    e^{-\actInt^+-(\actInt^+)^\dagger} \bullet\,
    \rangle_{\tensor[^\lMilne]{\CFT}{},\tensor[^\rMilne]{\CFT}{}}
    \;.
    \label{eq:entanglementAsInteraction}
\end{align}

With $\bullet$ in $\tensor[^\rMilne]{\CFT}{}$, the holographic dual of the
thermal expectation value
\eqref{eq:thermalState} arises by tracing out $\tensor[^\lMilne]{\CFT}{}$, \ie{} $\langle   e^{-{}^\rMilne\actInt} \,\bullet\,
    \rangle_{\tensor[^\rMilne]{\CFT}{}}$ with
\begin{align}
  &  {}^\rMilne\actInt 
    = -\int_0^\infty \frac{d\lambda\,
    e^{-2 \pi\lambda}}{(2\pi)^3}
    \frac{1+\lambda^2}{2\lambda}
    \!\int \celVolForm
    \tensor[^\rMilne]{\mathcal{O}}{^{1- i\lambda}}
    \cdot
    \tensor[^\rMilne]{\mathcal{O}}{^{1+ i\lambda}}\,.
    \label{eq:nice}
\end{align}
Expanding the exponential in \cref{eq:nice} yields a series of correlation
functions weighted by $(e^{-2\pi\lambda})^n$ and the entanglement entropy
corresponds to the von Neumann entropy of
this distribution. This procedure is reminiscent of
\cite{Mollabashi:2014qfa}, which examines entanglement between two interacting
\glspl{CFT}. However, their framework allows for a standard evaluation of entanglement
entropy of the resulting mixed state, which is not the case for the \gls{CCFT}.

Our calculations relied on the Maxwell theory being both free and Weyl
invariant. So, what general lessons have been learned? Much of the analysis
would carry through for a (weakly) interacting conformal gauge theory (\eg{}
${\cal N}=4$ super-Yang-Mills). However, one clear distinction arises since the
unitary relation \eqref{eq:blocksUnitaryRelation} fails. Hence, there is not a
separation \labelcref{EE55} between independent hard and soft contributions to
the entanglement entropy. If Weyl invariance is also lost, \eg{} by introducing
massive particles, asymptotic states would still decompose in terms of modes
with support to the future and past of the cut on $\mathscr{I}^+$ and the
Minkowski vacuum $\ket{0}$ would be some entangled state of these modes. The
decomposition would again divide the \gls{CCFT} into two sectors interacting
through the entangling operator.

While one sector would still correspond to the $\tensor[^\rMilne]{\CFT}{}$ dual
to the $\rMilne$ patch, it is interesting to speculate on the organization of
the remaining modes. Recall that the full Minkowski spacetime can be foliated
with (Euclidean) \gls{AdS} and dS slices \cite{deBoer:2003vf,Cheung:2016iub,
  Ogawa:2022fhy, Iacobacci:2022yjo,Sleight:2023ojm}, as in
\cref{fig:bulkSubregions}. One might conjecture that the
  $P,Q,R$ patches have their own dual CFTs that mutually interact to exchange
excitations and encode entanglement in spacetime. It would be interesting to
explore these speculations further.


Further details can be found in \cite{Chen:2024kuq}. There, we evaluate
the edge mode partition function \cite{Donnelly:2015hxa,Donnelly:2014fua}, from
which $S_{\rm Sh}(p)$ is easily computed. 
We also draw connections to the \gls{CFT} renormalization and cutoff scales. \\


\section*{Acknowledgements}

We would like to thank Freddy Cachazo, Laurent Freidel and especially Sabrina
Pasterski for helpful discussions. Research at Perimeter Institute is supported
in part by the Government of Canada through the Department of Innovation,
Science, and Economic Development Canada and by the Province of Ontario through
the Ministry of Colleges and Universities. HZC and RCM were also supported by the Natural
Sciences and Engineering Research Council of Canada. RCM was also supported by funding from the
BMO Financial Group and the Simons Foundation ``It from Qubit''
collaboration. AR is supported by the Heising-Simons Foundation ``Observational
Signatures of Quantum Gravity'' collaboration grant 2021-2817 and acknowledges
Perimeter Institute for hospitality while this work has been completed.%

\section*{Appendix: Entanglement from path integrals}

\begin{figure}[t!]
\vspace{10pt}
    \sidesubfloat[]{
    \includegraphics[scale=0.7]{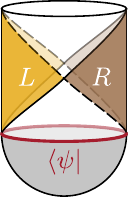}
    \label{fig:pathIntegralSphere}
  }  \hspace{25pt} \sidesubfloat[]{
    \includegraphics[scale=0.7]{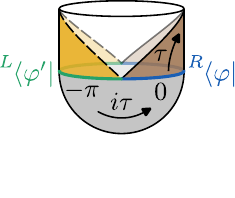}
    \label{fig:pathIntegralMilne}
  } 
  \caption{\protect\subref{fig:pathIntegralSphere} The
    common Einstein and Minkowski vacuum is prepared by a Euclidean path
    integral over an $S^4$ hemisphere. \protect\subref{fig:pathIntegralMilne}
    The same path integral prepares an $\lMilne,\rMilne$
      thermofield double.}
  \label{fig:pathIntegral}
\end{figure}
Weyl invariance of Maxwell theory allows us to prepare the vacuum
state $\ket{0}$ of the Minkowski and Einstein spacetimes with a Euclidean path
integral over an $S^4$ hemisphere, as shown in \cref{fig:pathIntegralSphere}.
However, the same path integral prepares an entangled thermal state from the Milne
perspective -- see \cref{fig:pathIntegralMilne}. 
 That is, the overlap
$\tensor[^\lMilne]{\bra{\varphi'}}{}\tensor[^\rMilne]{\braket{\varphi}{0}}{}$
with $\lMilne,\rMilne$ states coincides with the matrix element of an evolution
by $\pi$ in imaginary Milne time $\milnet$:
\begin{align}
  \tensor[^\lMilne]{\bra{\varphi'}}{}
  \tensor[^\rMilne]{\braket{\varphi}{0}}{}
  &\propto \tensor[^\rMilne]{\bra{\varphi}}{}
    e^{-\pi \tensor[^\rMilne]{H}{}}
    \tensor[^\rMilne]{\ket{\shad{\varphi}'}}{}
    \label{eq:vacuumMatElement}
  \\
  \implies
  \ket{0}
  &\propto \sum_i e^{-\pi E_i}
    \tensor[^\lMilne]{|\tilde{E}_i\rangle}{}
    \tensor[^\rMilne]{\ket{E_i}}{}
    \;,
    \label{eq:tfd}
\end{align}
where $\tensor[^\rMilne]{H}{}$ is the $\rMilne$ Hamiltonian, with eigenstates
$\tensor[^\rMilne]{\ket{E_i}}{}$ and eigenvalues $E_i$. A CPT transformation in
\cref{eq:vacuumMatElement} reinterprets the path integral boundary condition set
by the $\lMilne$ state $\tensor[^\lMilne]{\bra{\varphi'}}{}$ as an $\rMilne$
state $\tensor[^\rMilne]{\ket{\shad{\varphi}'}}{}$ -- for
Maxwell theory, this
coincides with the shadow transformation. Hence, $\ket{0}$ is a thermofield
double state on $\lMilne$ and $\rMilne$, and tracing over the $\lMilne$ Hilbert space
yields the thermal state
\begin{align}
  \tensor[^{\rMilne}]{\rho}{}[0]
  \equiv \tr_{\tensor[^L]{\mathcal{H}}{}}
  \ket{0}\bra{0}
  \propto e^{-2\pi \tensor[^\rMilne]{H}{}}.
  \label{eq:thermalState}
\end{align}
This closely parallels the path integral argument for the thermality of Rindler
wedges and, in fact, the $\lMilne,\rMilne$ Milne patches can be conformally
mapped to Rindler wedges.

As in the Rindler case, \cref{eq:tfd} may alternatively be deduced by
studying mode decompositions. With $\Delta=1+i\lambda$, \cref{eq:LR-dec}
coincides precisely with the Unruh \cite{Unruh:1976db} construction of positive-
and negative-definite Minkowski frequency modes in terms of $\lMilne,\rMilne$
modes with frequency $\lambda$. Thus, the resulting Bogoliubov transformations
relating Minkowski and Milne annihilation 
operators are identical to the Rindler
case. Demanding $\ket{0}$ and the Milne vacua
$\tensor[^{\lMilne/\rMilne}]{\ket{0}}{}$ be annihilated by the respective
annihilation operators, one finds
\begin{align}
  \ket{0}
  &\propto e^{- \actInt^+}
    \tensor[^{\lMilne}]{\ket{0}}{}
    \tensor[^{\rMilne}]{\ket{0}}{},
    \label{eq:relationBetweenVacua}
\end{align}
where the entangling operator is given by
\begin{align}
  \actInt^+
  &= -\int_0^\infty d\lambda\,
    e^{-\pi\lambda}
    \int \celVolForm\,
    (\tensor[^\lMilne]{\shad{a}}{}^{1+i\lambda})^\dagger
    \cdot
    (\tensor[^\rMilne]{a}{^{1-i\lambda}})^\dagger
    \label{eq:horse}
\end{align}
in terms of Milne creation operators. Here, $\celVolForm$ is the volume form
over celestial space \cite{Note1}. Since each Milne creation operator increments
the Milne energy $E$ by $\lambda$, we have recovered \cref{eq:tfd}.

\bibliography{biblio}




\pagebreak \widetext
\section*{Supplemetary material}

We begin with the standard metric on Minkowski spacetime
\begin{equation}
  \label{eq:Mink-metric}
  ds^2 = \eta_{\mu\nu}\, dX^{\mu} \,dX^{\nu}\,.
\end{equation}
The
  Cartesian coordinates are conveniently expressed as
\begin{align}
  X^\mu(u,r,{\bf z}) = u\, \hat{n}^\mu + r\, \hat{q}^\mu(\mathbf{z})
  \label{eq:X}
\end{align}
with
\begin{align}
  \label{eq:q}
  \hat{q}({\bf z})
  &= \frac{1}{2}(1 + z\bz, z + \bz, -i(z - \bz), 1 - z\bz)
    \;,
  \\
  \hat{n}^\mu
  &= \frac{1}{2}\Box^{(2)} \hat{q}^\mu
    = (1,0,0,-1) \;.
\end{align}
The metric \eqref{eq:Mink-metric} then becomes
\begin{equation}
  \label{eq:global-Mink}
  ds^2 = - 2\, du\, dr + r^2\,dz\,d\bar{z}\,,
\end{equation}
where $u, r \in (-\infty, \infty)$, and $(z,\bar{z})$ are a complex coordinates
on the celestial space.

 It is useful to define the metric $\celMet_{ab}$ on the
celestial space as
\begin{align}
  {\rm d}s_{\celPlane}^2({\bf z})
  = \celMet_{ab}\, {\rm d}z^{a}\, {\rm d}z^{b} \;, \quad  \celMet_{zz}
  = \celMet_{\bar{z}\bar{z} } = 0 \;,\quad
  \celMet_{z\bar{z}}
  = \celMet_{\bar{z} z}
  = \frac{1}{2} \;.
  \label{eq:yogurt}
\end{align}

The future Milne patch is covered by the coordinates $(\tau, \sigma, z,
\bar{z})$ related to the positive ranges of $u$ and $r$ in \cref{eq:global-Mink}
by
\begin{equation}
  u = \frac{e^{\tau - \sigma}}{2}\,, \qquad r = e^{\tau + \sigma}\,.
\end{equation}
In these coordinates, the metric in the future Milne patch takes the form
\begin{equation}
  ds^2_{\rm Milne} = e^{2\tau}\left(-d\tau^2 + d\sigma^2 + e^{2\sigma} dz d\bz \right) \,.
\end{equation}
Of course, we here recognize $ds^2_{\rm EAdS} = d\sigma^2 + e^{2\sigma} dz d\bz$
as the line element on a three-dimensional Euclidean \gls{AdS} geometry.


For the choice of {celestial} metric in \cref{eq:yogurt}, the covariant
derivative $D$ reduces to partial derivatives in complex coordinates
\begin{align}
  D_{z}
  &= \partial_{z} \;,\quad
    D_{\bar{z}}
    = \partial_{\bar{z}} \;
\end{align}
and the associated Laplacian is given by
\begin{align}
  \Box^{(2)}
  &\equiv
    D^a D_a
    = \celMet^{a b} D_a D_b
    = 4 \partial_{z} \partial_{\bar{z}} \;.
\end{align}
We will denote the celestial volume form and associated $\delta$-function
distribution by $\celVolForm$ and $\celDeltaFunc({\bfz},{\bf z}')$. These
satisfy
\begin{align}
  \int \celVolForm({\bf z}') \delta({\bf z},{\bf z}') f({\bf z}')
  &= f({\bfz}) \;.
\end{align}
In complex coordinates, they read
\begin{align}
  \celVolForm
  &= \frac{i}{2}
    {\rm d}z \wedge {\rm d} \bar{z} \;,\quad
    \celDeltaFunc({\bf z},{\bf z}')
    = -2i\, \delta^{(2)}(z-z',\bar{z}-\bar{z}') \;.
\end{align}
The celestial Green's function $\celG$ is defined by
\begin{align}
  \Box^{(2)}_{\bf z} \celG({\bf z},{\bf z}')
  &= \Box_{\bf z'}^{(2)} \celG({\bf z},{\bf z}')
    = \celDeltaFunc({\bf z},{\bf z}') \;,
    \label{eq:grape}
\end{align}
and, in complex coordinates, is given by
\begin{align}
  \celG({\bf z},{\bf z}')
  &= \frac{1}{4\pi}\log[(z-z')(\bar{z}-\bar{z}')] \;.
    \label{eq:celG}
\end{align}
Formally, $\celG$ acts as the inverse of $\celLap$ with the convolution:
\begin{align}
  [(\celLap)^{-1}f](\mathbf{w}) 
  = \int
  \celVolForm(\mathbf{z})\, G(\mathbf{w},\mathbf{z})\, f(\mathbf{z})
  \;.
  \label{eq:inverseLaplacian}
\end{align}

The frame fields $m^{\pm}_{a;\mu}$ in the \glspl{CPW}
\begin{align}
  \label{eq:spin-1-cpw-sm}
  A^{\Delta, \pm}_{a; \mu}({\bf w}; X)
  &=  \frac{
    m^{\pm}_{a;\mu}({\bf w}; X)
    }{
    [-\nullUnit(\mathbf{w})\cdot \minkX_{\pm}]^{\Delta}},
  &
    \minkX_{\pm}^{\mu}
  &= \minkX^{\mu} \mp i\reg \hat{n}^{\mu}
    \;,
\end{align}
 are given by \cite{Pasterski:2020pdk}
\begin{align}
  \label{eq:frame}
  m^{\pm}_{a;\mu}({\bf w}; X)
  &= \polar_{a; \mu}
    + \frac{\polar_{a} \cdot \minkX_{\pm}}{-\nullUnit\cdot \minkX_{\pm}}
    \nullUnit_{\mu}\;,
\end{align}
where $a$ is an index in the celestial space and labels the photon polarization
vector $\polar_{a}^\mu=\partial_a \hat{q}^\mu$.

The \glspl{CPW} \eqref{eq:spin-1-cpw-sm} can be decomposed into
a term involving the Mellin transform of a plane wave and a pure gauge piece
\cite{Donnay:2018neh}
\begin{align}
  A^{\Delta,\pm}({\bf w}; \minkX)
  &= \mathcal{K}^{\Delta,\pm} \, V^{\Delta,\pm}({\bf w};\minkX)
    + {\rm d}_X \alpha^{\Delta, \pm}({\bf w}; \minkX) \;,\quad 
    \mathcal{K}^{\Delta,\pm}
    = \frac{(\pm i)^\Delta(\Delta-1)}{\Gamma(\Delta+1)} \;,
    \label{eq:cake}
\end{align}
where
\begin{align}
  V^{\Delta,\pm}_{a;\mu}({\bf w};\minkX)
  =\polar_{a;\mu}
  \int_0^{\infty} {\rm d}\celFreq\, \celFreq^{\Delta - 1} e^{\pm i\celFreq \nullUnit\cdot \minkX_{\pm}} \;, \quad  \alpha^{\Delta,\pm}_a({\bf w}; \minkX)
  &= \frac{1}{\Delta} \frac{\polar_a \cdot \minkX_{\pm}}{(-\nullUnit \cdot \minkX_{\pm})^{\Delta}}.
    \label{eq:Mellin1}
\end{align}
Using the Klein-Gordon inner product of scalar plane waves
\cite{Pasterski:2017kqt, Donnay:2018neh} and the vanishing of the inner products
involving the pure gauge piece, one finds the inner product of \glspl{CPW}
\begin{align}
  \label{eq:cpw-ip}
  \langle
  A^{1+i\lambda,\pm}_a({\bf w}),
  A^{1+i\lambda',\pm}_{\bar{b}}({\bf w}')
  \rangle
  \eqIntegrated \pm 2(2\pi)^4
  \frac{\lambda \sinh(\pi \lambda) e^{\mp \pi \lambda}}{\pi(1 + \lambda^2)}
  \delta(\lambda - \lambda') \,
  \celMet_{ab}\celDeltaFunc({\bf w},{\bf w}').
\end{align}

 The shadow transformation yields an alternate set of \glspl{CPW}
\begin{align}
  \label{eq:shadow-spin-1-cpw}
  \tilde{A}^{\Delta, \pm}_a = \widetilde{A_a^{2-\Delta, \pm}} = (-X_{\pm}^2)^{\Delta-1}  A^{\Delta,\pm}_a
  \;.
\end{align}
  Here, $\tilde{A}^{\Delta, \pm}_a$ denotes the shadow mode with boost weight
  $\Delta$ while ${\widetilde{A_a^{\Delta, \pm}}}$ denotes the shadow
  transformation of the \gls{CPW} with weight $\Delta$.

 The $\Delta = 1$ ($\lambda=0$) modes give rise to the Goldstone
and conformally soft wavefunctions,
\begin{align}
  A^{\gold}_a({\bf w}, X)
  &\equiv
    \frac{A^{1,+}_a({\bf w}; X) + A^{1,-}_a({\bf w}; X)}{2}
    = {\rm d}_X \alpha^{\gold}_a,
    \nonumber\\
  A^{\cs}_a({\bf w}, X) &\equiv \frac{A^{\log,+}_a({\bf w}, X) - A^{\log,-}_a({\bf w}, X)}{2\pi i},
                          \label{eq:soft-wf}  
\end{align}
where $\alpha^\gold_a({\bf w}, X)$ parameterizes an asymptotic symmetry
transformation, and 
\begin{align}
  A^{\log,\pm}_a
  \equiv \lim_{\Delta \rightarrow 1} \partial_{\Delta}\left[
  A^{\Delta,\pm}_a + \tilde{A}^{2 - \Delta, \pm}_a
  \right] = -\log(-X_\pm^2)  A^{1,\pm}_a
  .
  \nonumber
\end{align}
$A^{\gold}$ and $A^{\cs}$ are canonically conjugate with respect to the standard
conserved inner product \cite{Donnay:2018neh}.

The non-integer $\Delta$ \glspl{CPW} \eqref{eq:spin-1-cpw-sm},
as well as the logarithmic constituents of the conformally soft modes
\eqref{eq:soft-wf} have branch-cuts. We choose branches of logarithms and
non-integer powers such that, at finite $\reg$, functions do not cross branch
cuts as $X$ varies through spacetime. In the $\reg\to 0$ limit, we have
\begin{align}
  \arg(-\nullUnit\cdot\minkX_\pm)
  &\eqReg \mp\pi \,\stepFunc(\nullUnit\cdot\minkX) \,,
  \\
  \arg(-\minkX_{\pm}^2)
  &\eqReg
    \begin{cases}
      0 & \text{$\minkX^2<0$ and $\minkX^0>0$} \\
      \mp \pi & \text{$\minkX^2>0$} \\
      \mp 2\pi & \text{$\minkX^2<0$ and $\minkX^0<0$}
    \end{cases} \;.
\end{align}

The asymptotic parameter in \cref{eq:soft-wf} admits the decomposition
\begin{align}
  \alpha^{\gold}_a({\bf w},\minkX)
  &\equiv \frac{\alpha^{1,+}_a({\bf w},\minkX) + \alpha^{1,-}_a({\bf w},\minkX)}{2},\\
  \alpha^{1,\pm}({\bf w};\minkX)
  &= -{\rm d}_{{\bf w}} \log(-\nullUnit\cdot \minkX_\pm)
    = - 4\pi {\rm d}_{{\bf w}} \celG({\bf z},{\bf w}) \left[
    1 + \frac{2(\minku\mp i\reg)}{\minkr|\celwh-\minkzh|^2}
    \right]^{-1}.
\end{align}
At large-$r$ these have the following expansions
\begin{align}
  \alpha^{1,\pm}({\bf w};\minkX)
  &= - 4\pi {\rm d}_{\bf w} \celG({\bf w},{\bf z})
    + \order{\minkr^{-1}\log\minkr}
    =
    \alpha^{\gold}({\bf w};\minkX).
    \label{eq:tea}
\end{align}
As explained in the main text, the  conformal primary wavefunctions \eqref{eq:cake} admit decompositions into left and right Milne constituents obeying
\begin{align}
  - &\langle
      \tensor[^{\lMilne}]{\tilde{A}}{}^{1 + i\lambda}_a({\bf w}),
      \tensor[^{\lMilne}]{\tilde{A}}{}^{1 + i\lambda'}_{\bar{b}}({\bf w}')
      \rangle
      = \langle
      \tensor[^{\rMilne}]{A}{}_a^{1 + i\lambda}({\bf w}),
      \tensor[^{\rMilne}]{A}{}_{\bar{b}}^{1 + i\lambda'}({\bf w}')
      \rangle
       \eqIntegrated (2\pi)^4 \frac{\lambda}{\pi(1+\lambda^2)}
      \,\delta(\lambda - \lambda')
      \,\gamma_{a b} \,\delta({\bf w},{\bf w}')\;,
      \label{eq:limitMinkProdRRLL}
\end{align}
where $\gamma_{a b}$ is the celestial space metric. The inner product of
$\lMilne$ and $\rMilne$ modes vanishes as $\reg\to 0$. Here, $\eqIntegrated$
denotes equality as $\epsilon \rightarrow 0$ as distributions in $\lambda$ when
the test functions are smooth near $\lambda \in \mathbb{R}.$


\end{document}